\RequirePackage{fix-cm}
\documentclass[twocolumn]{svjour3}          
\smartqed  
\usepackage{graphicx}
\usepackage[english]{babel}
\usepackage[utf8]{inputenc}
\usepackage{amsmath}
\usepackage{caption}
\usepackage{subcaption}
\usepackage{hyperref}
\usepackage[colorinlistoftodos]{todonotes}
\DeclareUnicodeCharacter{00A0}{ }
\journalname{Nonlinear Dynamics}
\begin{document}

\title{Detecting Dynamical States from Noisy Time Series using Bicoherence}


\author{Sandip V. George         \and
        G. Ambika \and
	R. Misra}


\institute{Sandip V. George \and G. Ambika \at
	Indian Institute of Science Education and Research, Dr Homi Bhabha Road, Pune 411008, India\\
	\email{sandip.varkey@students.iiserpune.ac.in}\\
	\email{g.ambika@iiserpune.ac.in}
	\and R. Misra \at
	Inter University Centre for Astronomy and Astrophysics, Ganeshkhind, Pune 411 007, India\\
	\email{rmisra@iucaa.in}
	}

\date{Received: date / Accepted: date}
%

\maketitle

\begin{abstract}
Deriving meaningful information from observational data is often restricted by many limiting factors, the most important of which is the presence of noise. In this work, we present the use of the bicoherence function to extract information about the underlying nonlinearity from noisy time series. We show that a  system evolving in the presence of noise which has its dynamical state concealed from quantifiers like the power spectrum and correlation dimension D$_2$, can be revealed using the bicoherence function. We define an index called main peak bicoherence function as the bicoherence associated with the maximal power spectral peak. We show that this index is extremely useful while dealing with quasi-periodic data as it can distinguish strange nonchaotic behavior from quasi-periodicity even with added noise. We demonstrate this in a real world scenario, by taking the bicoherence of variable stars showing period doubling and strange nonchaotic behavior. Our results indicate that bicoherence analysis can also bypass the method of surrogate analysis using Fourier phase randomization, used to differentiate linear stochastic processes from nonlinear ones, in conventional methods involving measures like D$_2$.
\end{abstract}

\section{Introduction}
\label{sec:Sec1}
The presence of noise in real time series or data, is a problem that has been addressed frequently in recent times. Except in very limited circumstances, it is difficult to derive meaningful information about the underlying dynamics in such cases \cite{fra01}. Despite having no information about the underlying non-linearity in a times series, the power spectrum is used extensively in studying nonlinear dynamical systems. At the same time the use of polyspectra like bi and tri spectra to understand dynamical behavior is rather limited \cite{pez90,pez92,wan01,eva00}. Limitations posed by the power of the dominant nonlinearity and need for high computation speeds would perhaps partially explain the reason why the use of polyspectra has been less popular in analysis of dynamical systems.

Nonlinear dynamical systems can exhibit a rich variety of states like periodic, quasi-periodic, intermittently chaotic, strange nonchaotic and chaotic dynamics \cite{hil00,pra01}. The problem of studying these dynamical behaviors becomes more acute if the dynamical system is evolving in a noisy environment \cite{lon03,sch85,bas15}. The presence of noise also limits the useful information one can derive from observational data or time series and may cast doubts on the conclusions drawn from any standard analysis. This is especially true when trying to quantify nonlinearity in the system \cite{cas91,ser07}. Moreover, establishing non-linearity based on techniques like D$_2$ and Lyapunov exponents can be very difficult in the presence of noise, especially if it is colored \cite{han95,osb89,yan11}. The hypothesis testing using Fourier phase randomized surrogates often helps to rule out false positives but gives no further information about the underlying dynamical behavior. In this paper we advocate the use of bispectrum based techniques to analyze such time series to understand the underlying dynamical properties. 

Bispectral analysis of univariate time series is very popular due to the inherent advantages it offers. For instance it retains the phase information, unlike the power spectrum which is phase blind. Further it does not differentiate between the different kinds of mean zero symmetrically distributed noises, whether white or colored \cite{tot15}. Due to these inherent advantages it has been widely used in a variety of applications like in EEG analysis, damping identification, CMBR analysis, analysis of Chua circuit etc.  \cite{gaj98,haj00,fer09,elg93}. 

In this work, we define a quantity, the main peak bicoherence ($b_F(f)$), which is just the bicoherence function associated with the maximal frequency in the power spectrum. We initially show how all the peaks in the quasi-periodic and strange nonchaotic power spectra appear significantly in $b_F(f)$. We show how this function acts as a filter to pick out true peaks from a noisy time series either from dynamical systems or from observational data. This implies that Fourier phase randomized surrogate testing that is popularly used in conjunction with standard nonlinear quantifiers like correlation dimension and entropy to distinguish linear stochastic processes, can be bypassed when using bispectral analysis to detect nonlinearity.

We show in Section \ref{sec:sec2} that a dynamical system, like R{\"o}ssler, evolving in the presence of noise in the limit cycle region has a power spectrum similar to the chaotic noiseless R{\"o}ssler, and hence can give false conclusions. Also the doubly driven pendulum, with an additive colored noise process, in the quasi-periodic regime, can be misrepresented as being strange nonchaotic as shown in Section \ref{sec:sec3}. We illustrate how the bicoherence function can be used to unravel the actual dynamical behavior even in such cases.

In this context, as an important application to real world data, we consider variable stars in Section \ref{sec:sec4}, which is an area where modeling using a dynamical systems approach has been largely successful \cite{reg06}. Many nonlinear models have been developed, that describe the intensity variations in these stars \cite{bak79,ste86}. Nonlinear time series analysis techniques have been used to detect deterministic chaos in some of them \cite{buc95,geo15}. Recently availability of high quality light curves from the Kepler space telescope has led to heightened interest in variable star astrophysics, with fascinating results reported like the detection of period doubling in RRabc Lyrae stars and strange nonchaotic behavior in RRc Lyrae stars \cite{sza10,mos15,lin15}. We use the full bicoherence function and the main peak bicoherence function to analyze both these phenomena. Apart from testing our analysis in a real world setting, the bicoherence also reveals interesting features about the stars' dynamics that wouldn't be apparent from other analyzes.

The paper is organized as follows. In Section \ref{sec:sec2} we use the bicoherence function to analyze the noisy limit cycles of the R{\"o}ssler system. In Section \ref{sec:sec3} we define a main peak bicoherence function. We use it to filter the power spectral scaling behavior of a quasi-periodic time series with colored noise and a strange nonchaotic time series. In Section \ref{sec:sec4} we apply these techniques to analyze the light curves of variable stars. Section \ref{sec:sec4a} analyses light curves of stars in the period doubled state. Section \ref{sec:sec4b} analyses light curves of strange nonchaotic stars. Section \ref{sec:sec5} summarizes the contents of the paper and lists the advantages and disadvantages of the methods described. 

\section{Analysis of Noisy Limit Cycles}
\label{sec:sec2}
We start by considering how limit cycle oscillations in the presence of noise can be identified using bicoherence. For this we revisit the basic definitions.  The bispectrum is defined as the Fourier transform of the three-point auto correlation function. This can be equivalently written as
\begin{equation}
B(f_1,f_2)=\langle A(f_1) A(f_2) A^*(f_1+f_2) \rangle_\infty
\end{equation}
where A(f) is the Fourier transform of the signal at frequency, f and A$^*$(f) is its complex conjugate  \cite{tot15,rao12}
For k segments of the time series, the normalized bispectrum (or bicoherence) is defined as 
\begin{equation}
\label{eqn:e4}
b(f_1,f_2)=\frac{|\sum\limits_{i=1}^k A(f_1) A(f_2) A^*(f_1+f_2)|}{\sum\limits_{i=1}^k|A(f_1) A(f_2) A^*(f_1+f_2)|}
\end{equation}
It is clear that the bicoherence function takes values between zero and one \cite{hay07}. 

Due to the inherent symmetries, the bicoherence function of a real valued process can be completely defined in the triangular region given by \cite{tot15}
\begin{eqnarray}
f_2 > 0  \nonumber \\
f_1 \geq f_2 \nonumber \\
f_1+f_2 \leq f_{max} 
\end{eqnarray}
where $f_{max}$ is the Nyquist frequency (or half of the sampling rate) of the time series \cite{pre07}.

\begin{figure*}
\begin{subfigure}{\textwidth}
\includegraphics[width=\textwidth]{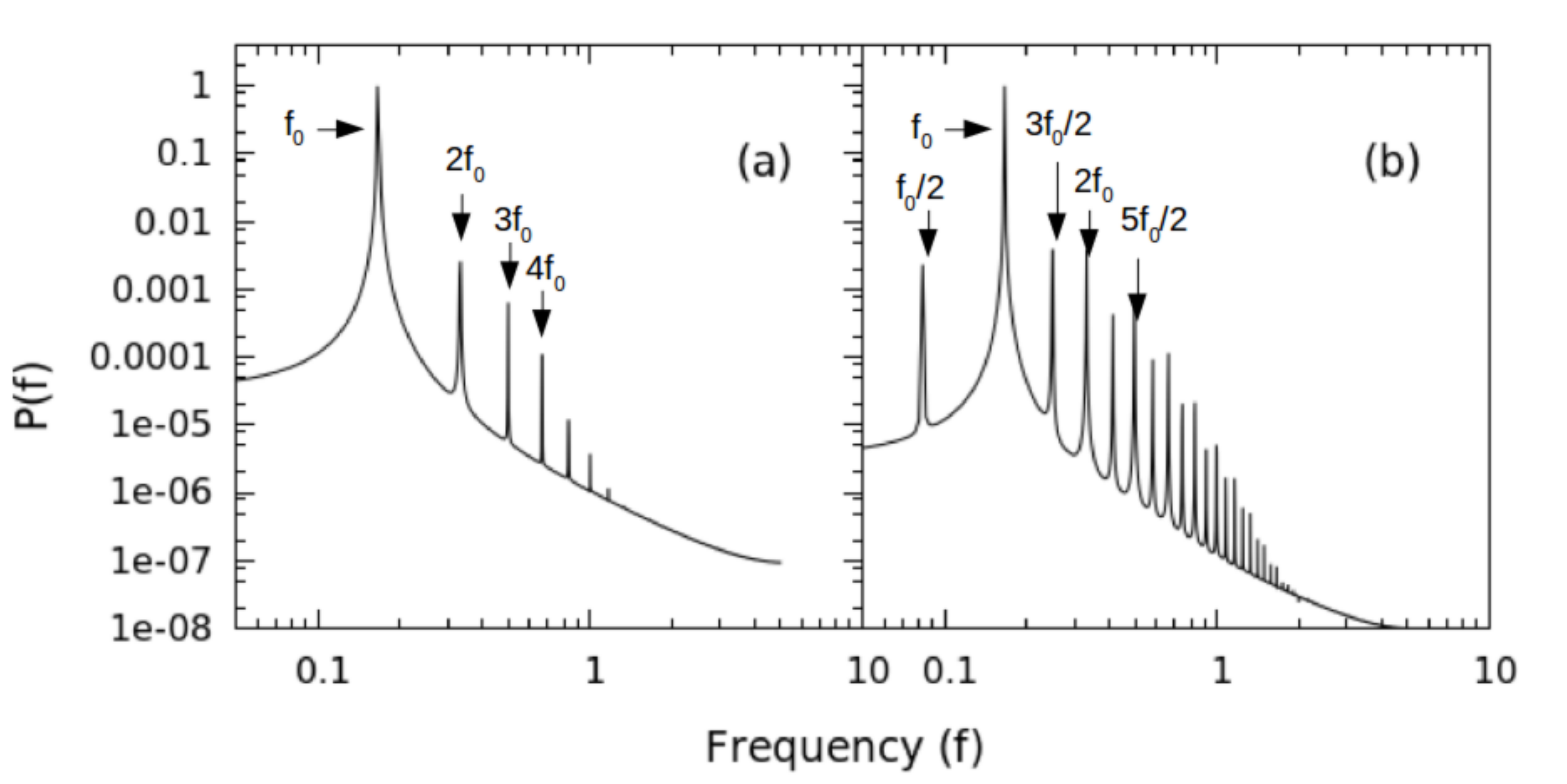}
\end{subfigure}
\begin{subfigure}{\textwidth}
\includegraphics[width=\textwidth]{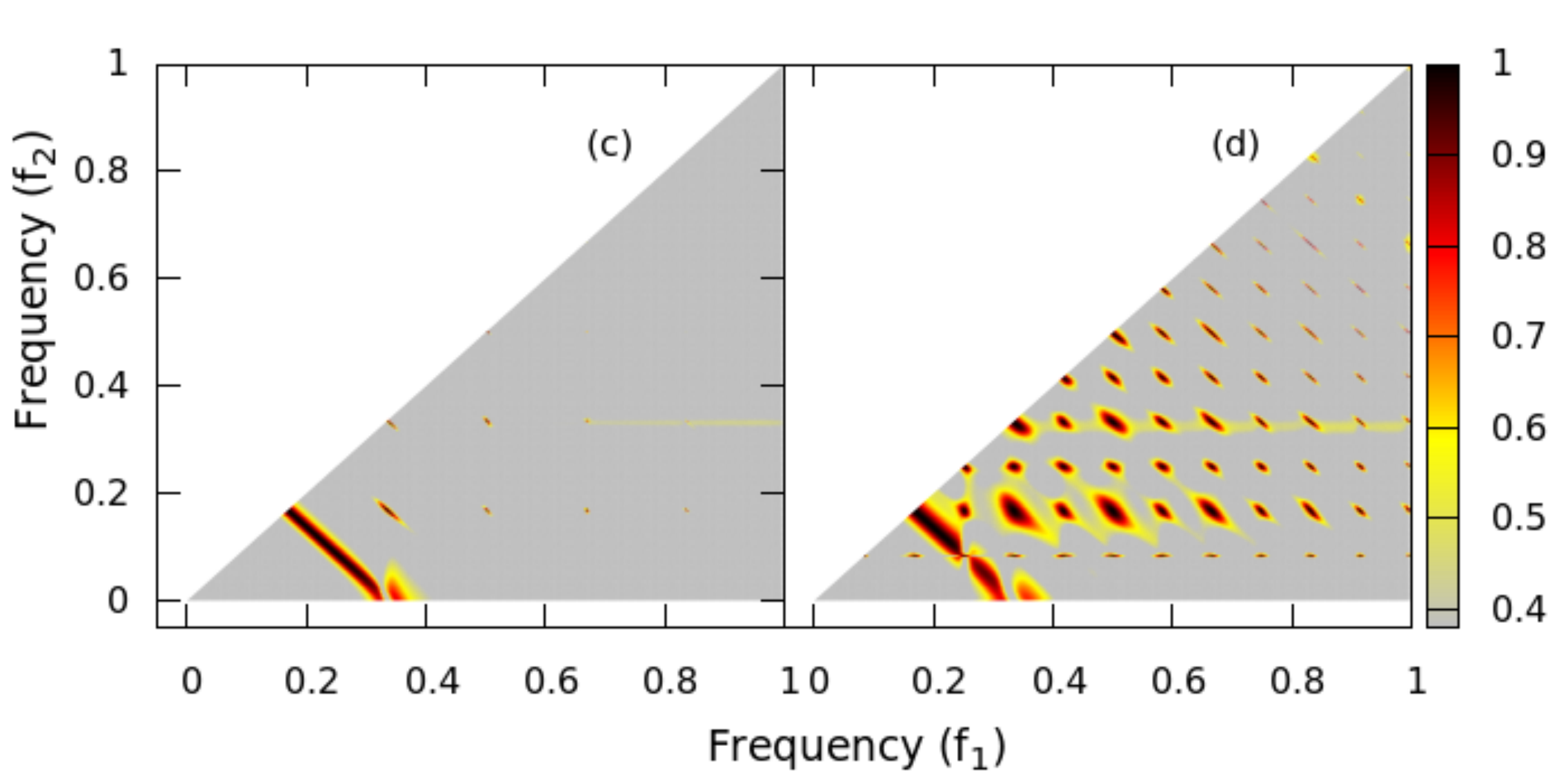}
\end{subfigure}
\caption{\label{fig:pdroute} Power spectra and the corresponding bicoherence graphs for the period 1 and period 2 states of the R{\"o}ssler system. Power spectra presented are (a) period-1 and (b) period-2. The primary frequency is at $f_0$=0.166 Hz in (a). The half harmonic can be seen at 0.083Hz in (b). The bicoherence graphs have been zoomed into the range [0,1], beyond which the power spectrum shows no interesting behavior. Significant bicoherence can be seen at the fundamental-harmonic and harmonic-harmonic frequency pairs in (c), while in (d) it can also be seen at the fundamental-half harmonic and harmonic-half harmonic frequency pairs.}
\end{figure*}

As a standard nonlinear dynamical system in the periodic limit cycle regime, we study the R{\"o}ssler equations evolving in the presence of noise. The equations are given by
\begin{equation}
\begin{aligned}
\dot{x}&=-y-z+\eta(t)\\
\dot{y}&=x+ay \\
\dot{z}&=b+z(x-c)\\
\end{aligned}
\end{equation}
where a=0.1, b=0.1, c is varied for the regime under consideration and $\eta$(t) is $\delta$-correlated mean zero Gaussian white noise with the standard deviation set at 2.0 for periodic region(c=4, period 1 and c=6.2, period 2) and at 5.0 for chaotic regime (c=18). By integrating the system using the fourth order Runge Kutta algorithm for 2$^{17}$ time steps, we compute the power spectrum of the time series of the x variable for all the cases, by dividing it into 32 segments. For comparison, we also consider the power spectrum for period 1 and period 2 limit cycles of R{\"o}ssler in the absence of noise in Figures \ref{fig:pdroute} a and b. For this case bicoherence planes have already been reported earlier in literature but added here for completion in Figures \ref{fig:pdroute} c and d \cite{pez90}. The islands of high bicoherence and straight lines are artifacts of finite widths of peaks because of the finite size of data. The bicoherence is only plotted for values above a 99$\%$ significance level, given by $\sqrt{\frac{9.2}{dof}}$  \cite{cha93}.

\begin{figure*}
\centering
\includegraphics[width=\textwidth]{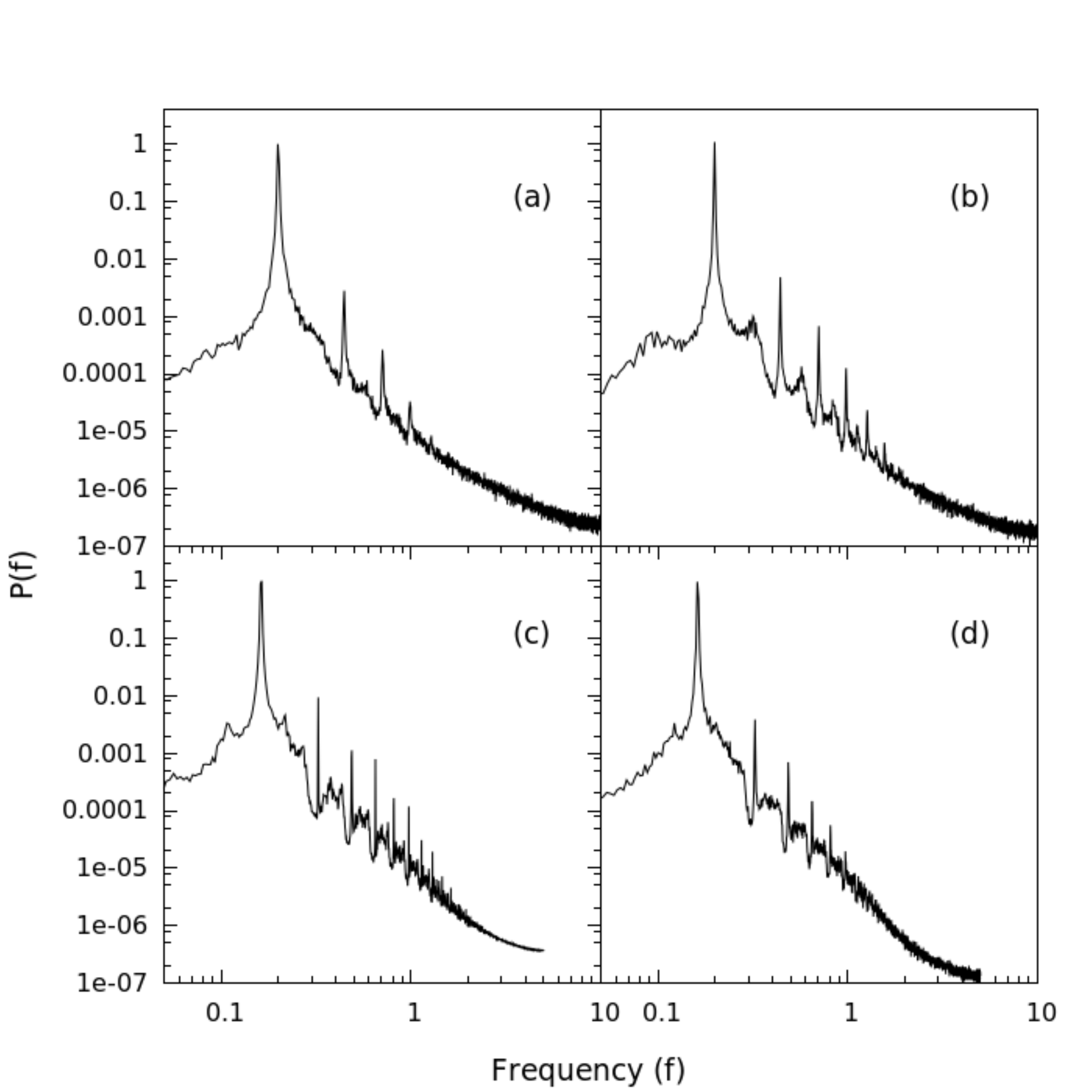}
\caption{\label{fig:PS_pdrn} Power spectra of R{\"o}ssler system for (a)period 1 regime with noise (b)period 2 regime with noise (c) chaotic regime without noise  and (d) chaotic regime with noise. Without a priori knowledge of the system, from the power spectrum alone, it becomes difficult to identify whether the interpeak continuum is due to dynamics or noise, from the power spectrum alone.}
\end{figure*}

Unlike the power spectra for the noiseless R{\"o}ssler limit cycles in Figure \ref{fig:pdroute}, the R{\"o}ssler limit cycles evolved in the presence of noise, shows a continuum inter-peak power in Figures \ref{fig:PS_pdrn}a and b. This continuum inter-peak power is the signature of chaotic behavior, as is evident from the spectrum of the noise free chaotic R{\"o}ssler shown in Figure \ref{fig:PS_pdrn}c.  Further the fine features of the power spectrum of chaotic R{\"o}ssler is lost when evolved with background noise as can be seen from Figure \ref{fig:PS_pdrn}d. In this case, the power spectrum becomes increasingly similar to the noisy limit cycle power spectrum. Inferring whether the continuum inter-peak power is due to noise or chaos or a combination of both, is essential to deciphering the underlying dynamical behavior exhibited by the system and hence an important step in modeling it. We point out that without an a priori knowledge of the system, this is not apparent from the power spectrum alone. 

\begin{figure*}
\centering
\includegraphics[width=\textwidth]{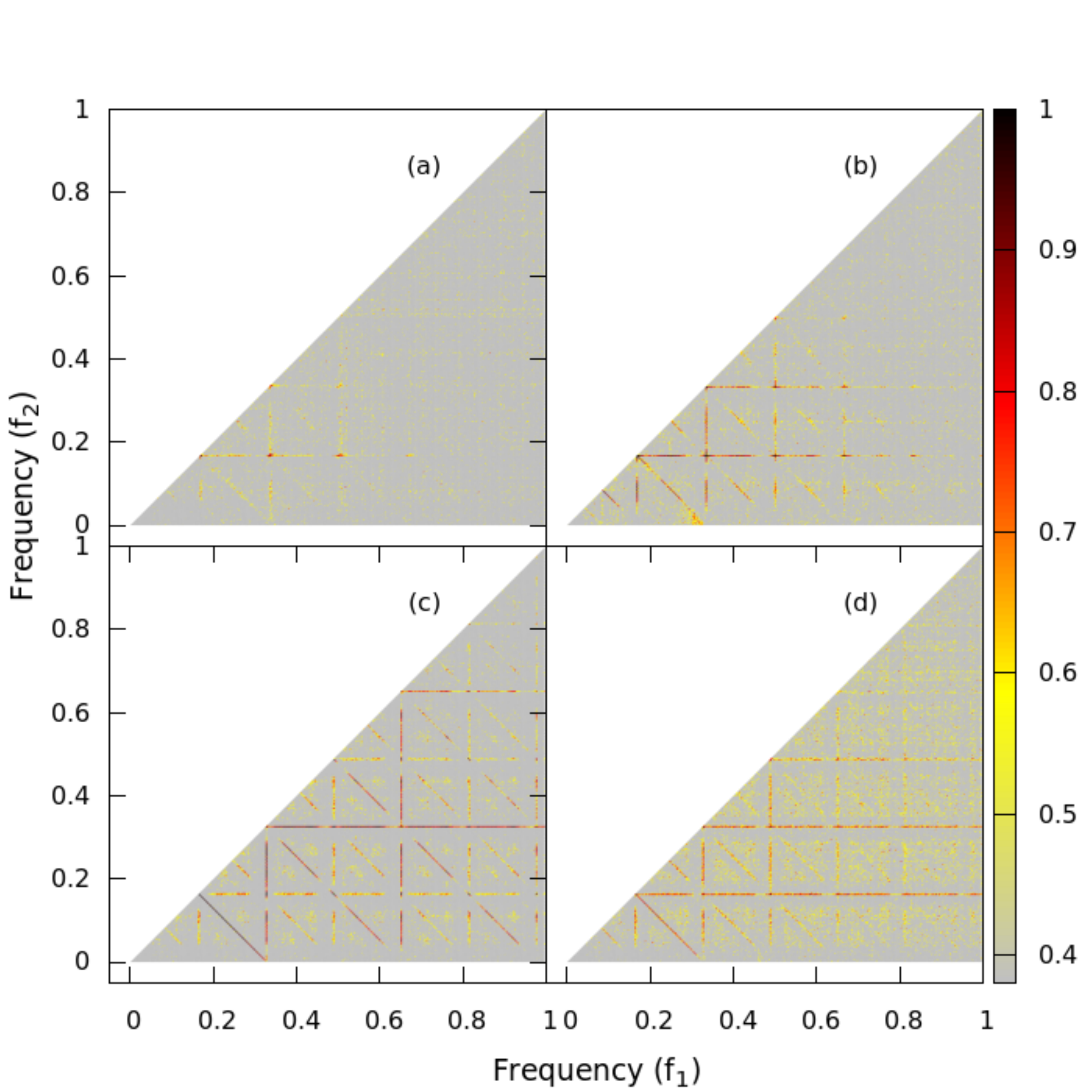}
\caption{\label{fig:BS_pdrn} Bicoherence of R{\"o}ssler system corresponding to the states shown in Figure 2. (a)period 1 regime with noise (b)period 2 regime with noise (c) chaotic regime without noise  and (d) chaotic regime with noise. Even when evolved in a noisy environment, the broad features of chaos are retained in bicoherence in (d).}
\end{figure*}

We present the bicoherence for all the above cases in Figure \ref{fig:BS_pdrn}. As in Figure \ref{fig:pdroute} the relevant part of the bicoherence plane is only up to around 1 Hz. Hence we zoom into this region in Figure \ref{fig:BS_pdrn}. The bicoherence can see through the noise contamination that affects standard quantifiers and significant values of bicoherence can be seen only for the primary frequency and its harmonics in Figures \ref{fig:BS_pdrn}a and b. The underlying limit cycle behavior can hence be clearly identified. The noiseless chaotic behavior remains distinctly different from the noisy limit cycle behavior in Figure \ref{fig:BS_pdrn}. Moreover, the background noise contamination in the chaotic R{\"o}ssler leaves the bicoherence function broadly unaffected.
\section{Main Peak Bicoherence Function and Quasi-periodicity}
\label{sec:sec3}
Quasi-periodicity occurs in dynamical systems when there are two incommensurate frequencies $F_1$ and $F_2$ in the system. Then peaks at $F_1$ and $F_2$ and all the linear combinations of $F_1$ and $F_2$ are present in the corresponding power spectrum  \cite{lak12}. From the definition of the bicoherence function (equation \ref{eqn:e4}), we see that these peaks would result in a significant bicoherence at ($F_1$,$F_2$) and ($F_1-F_2$,$F_2$). Since all the major peaks in this case would come from the linear combinations of the two main frequencies, we introduce an index called main peak bicoherence function ($b_F(f)$) defined as
\begin{equation}
b_{F}(f)=\frac{|\sum\limits_{i=1}^k A(F) A(f) A^*(F+f)|}{\sum\limits_{i=1}^k|A(F) A(f) A^*(F+f)|}
\end{equation}
where $F$ is the maximal peak in the power spectrum. 
Such quasi-periodic forcing in nonlinear systems can also lead to strange nonchaotic behavior \cite{pra01}. Systems exhibiting strange nonchaotic behavior have an attractor with fractal geometry but no exponential divergence of nearby trajectories, giving the peculiar combination of a non integer fractal dimension but a non positive maximal Lyapunov exponent. As mentioned earlier, in real data, effects like noise affect both these quantifiers \cite{han95,har09}.

To demonstrate both quasi-periodic and strange nonchaotic behaviors, we consider the standard system of a quasi-periodically forced pendulum, whose equations are given by,
\begin{equation}
\frac{1}{p}\frac{d^2\phi}{dt^2}+\frac{d\phi}{dt}-cos\phi=g(t) 
\label{eqn:e8}
\end{equation}

where 
\begin{equation}
g(t)=K+V[cos(\Omega_1t)+cos(\Omega_2t)] \nonumber
\end{equation}

\begin{figure*}
\centering
\begin{subfigure}{\textwidth}
\includegraphics[width=\textwidth]{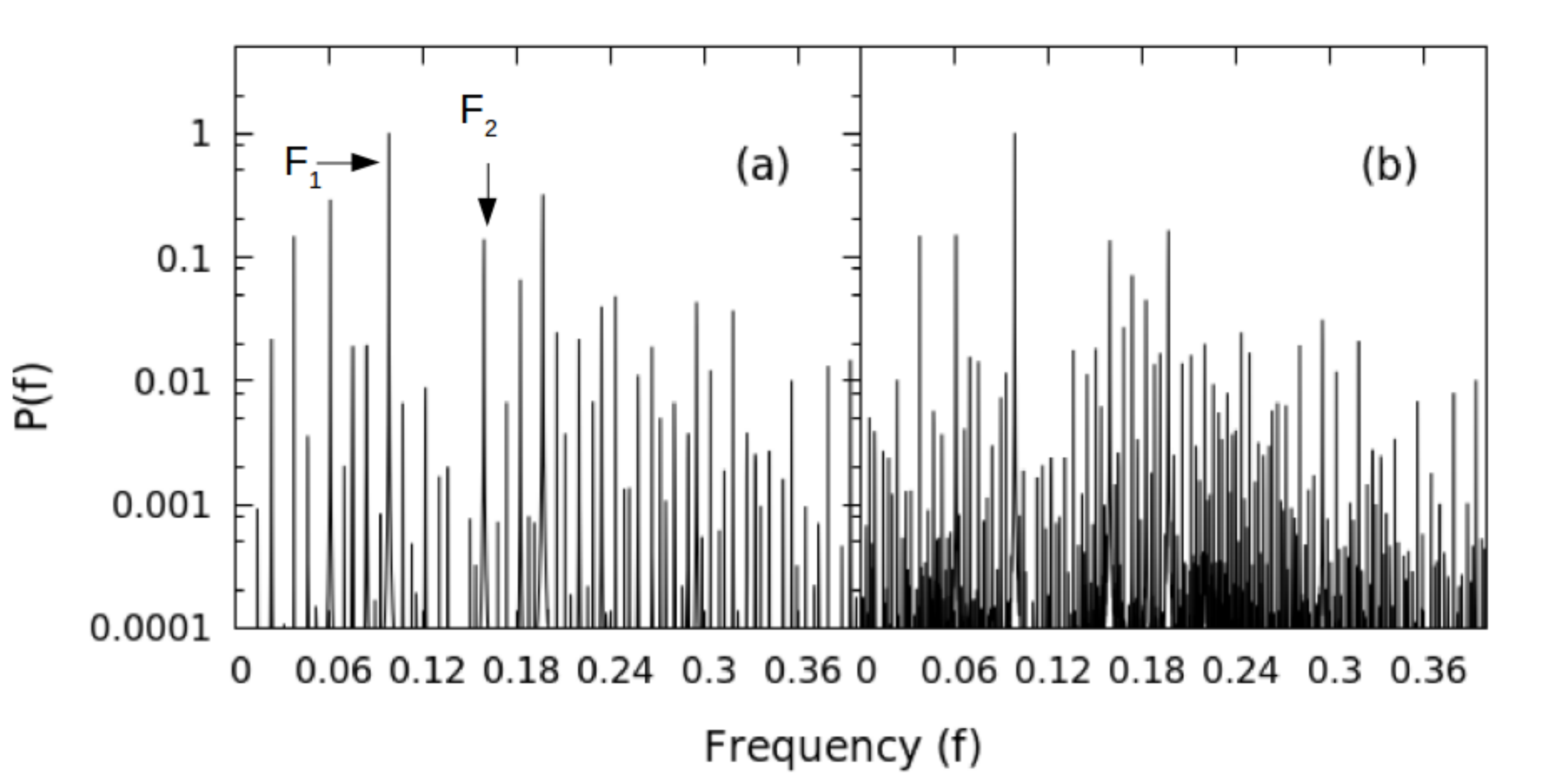}

\end{subfigure}

\begin{subfigure}{\textwidth}
\includegraphics[width=\textwidth]{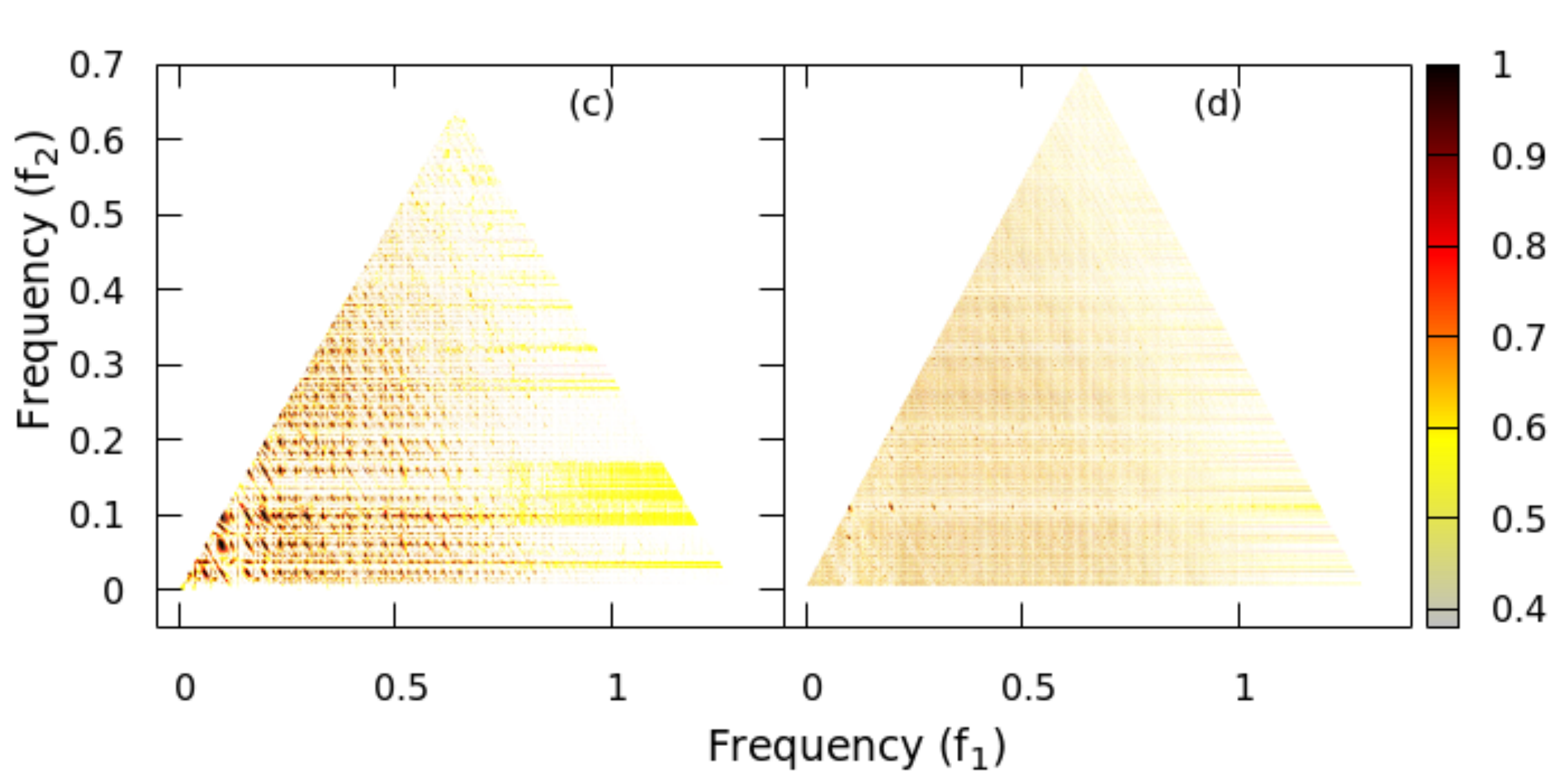}

\end{subfigure}

\caption{\label{fig:qpsnc} Power spectra for (a)quasi-periodic and (b)strange nonchaotic data from a doubly driven pendulum. The forcing frequencies are at $F_1$=0.10 Hz and $F_2$=0.16 Hz. Full bicoherence planes for (c) quasi-periodic state (d) strange nonchaotic state. (c) has islands of high bicoherence as expected from a discrete power spectrum while in (d) a more uniform bicoherence spread over many frequency pairs can be seen.}
\end{figure*}

\begin{figure*}
\begin{subfigure}{\textwidth}
\includegraphics[width=\textwidth]{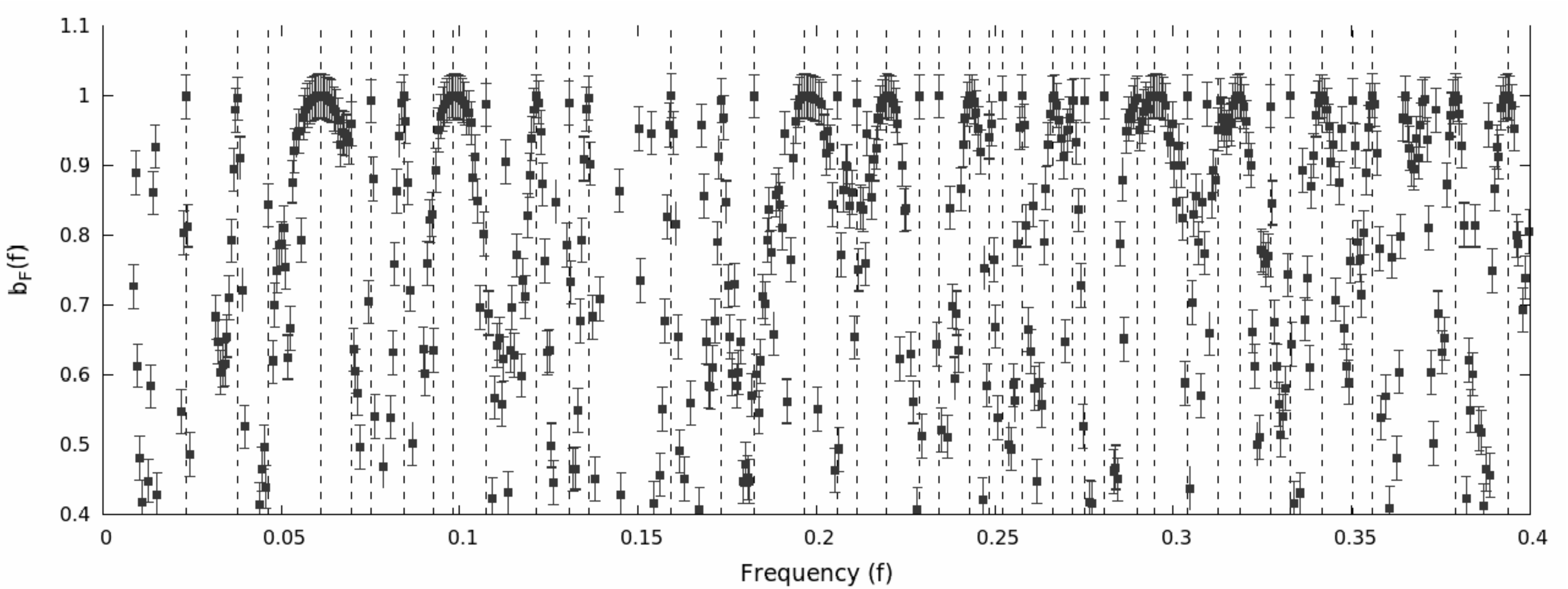}
\caption{\label{fig:mbcqp}}
\end{subfigure}
\begin{subfigure}{\textwidth}
\includegraphics[width=\textwidth]{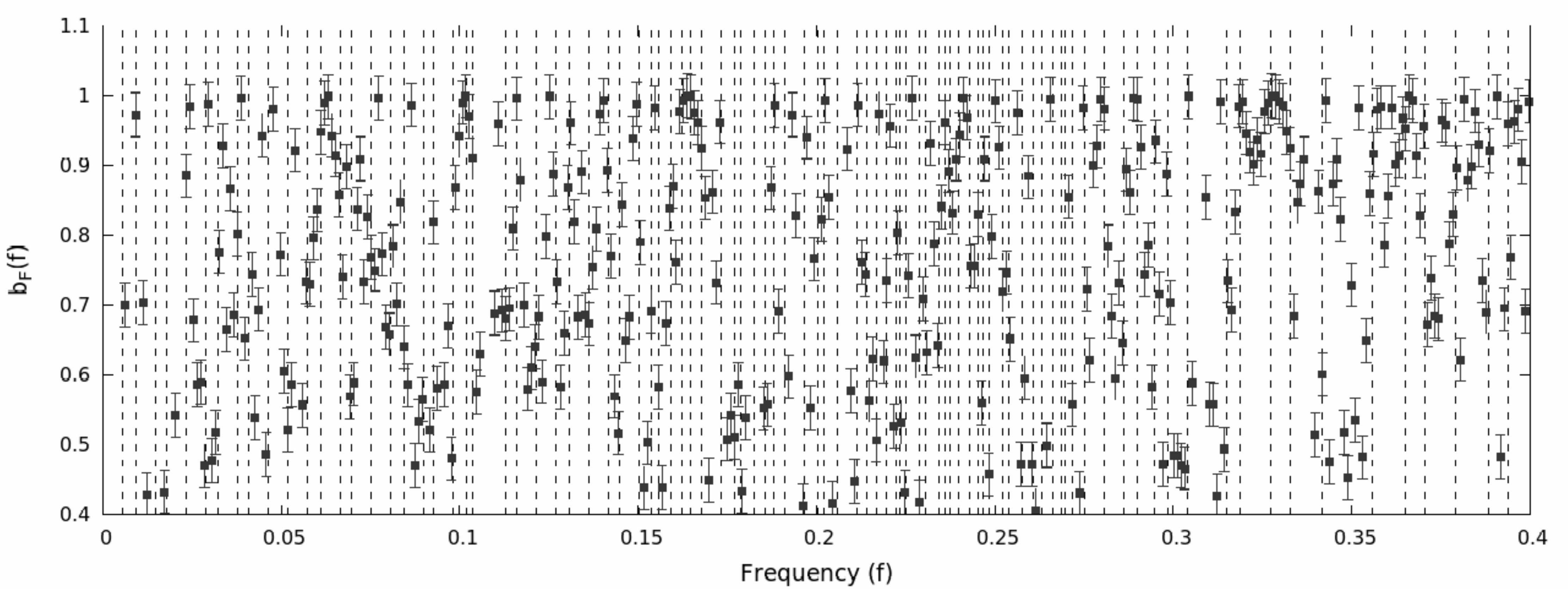}
\caption{\label{fig:mbcsnc}}
\end{subfigure}
\caption{\label{fig:mbcqpsnc} Main peak bicoherence ($b_F(f)$) vs frequency graphs for (a) quasi-periodic and (b) strange nonchaotic pendulum. The dashed vertical lines correspond to peaks above .001 in the power spectrum. All the peaks have significant bicoherence values. The bicoherence peaks that do not coincide with the vertical lines are either from true peaks below .001 or due to the finite peak widths.}
\end{figure*}

In this system quasi-periodicity is achieved for values $K$=1.34, $V$=0.55, $p$=3.0, $\Omega_1$=2$\pi F_1$=$\frac{\sqrt{5}-1}{2}$ and $\Omega_2$ =2$\pi F_2$ =1.0 , and strange non chaotic behavior is achieved by changing $K$ to 1.33 \cite{rom87}. We integrate the system for $2^{19}$ time-steps in both regimes. The computed power spectrum of the time series of the $y$ variable signal(where $y=\frac{d^2(\phi)}{dt^2}$) and its full bicoherence graphs in the quasi-periodic and strange nonchaotic regimes, calculated for 32 segments of 16384 points, are shown in Figure \ref{fig:qpsnc}. The two dynamical states are easily distinguishable both in power spectrum and bicoherence. We plot the corresponding $b_F(f)$ vs $f$ in Figure \ref{fig:mbcqp}. All the main peaks in the power spectrum also appear in $b_F(f)$. The vertical lines correspond to main peaks in the power spectrum (above .001). Figure \ref{fig:mbcsnc} also shows that all the main peaks in the strange nonchaotic power spectrum also have significant bicoherence. The standard deviation on the bicoherence estimated from $N$ different Fourier transforms is $\frac{1}{N}$ \cite{mac02}.


\subsection{Spectral scaling with bicoherence filter}
\label{sec:sec3a}
One of the more popular methods to identify strange nonchaotic behavior in real data has been the spectral scaling of peaks in the power spectrum \cite{lin15}. This is done by resampling the time series at a time step equal to its primary period (strobing), followed by calculating its power spectrum (which is equivalent to taking a Poincare section of the data). The threshold power is then increased in steps and the number of peaks above that threshold is counted, in the power spectrum. This scaling of this number is shown to follow a power law, if the system exhibits strange nonchaotic behavior \cite{hea91}. We find however, that if we have a quasi-periodically forced system, that has an added colored noise process, the spectral scaling of peaks in the power spectrum of the Poincare section in the quasi-periodic region is similar to the scaling exhibited by the noise-free strange nonchaotic power spectrum.

\begin{figure*}
\centering
\includegraphics[width=\textwidth]{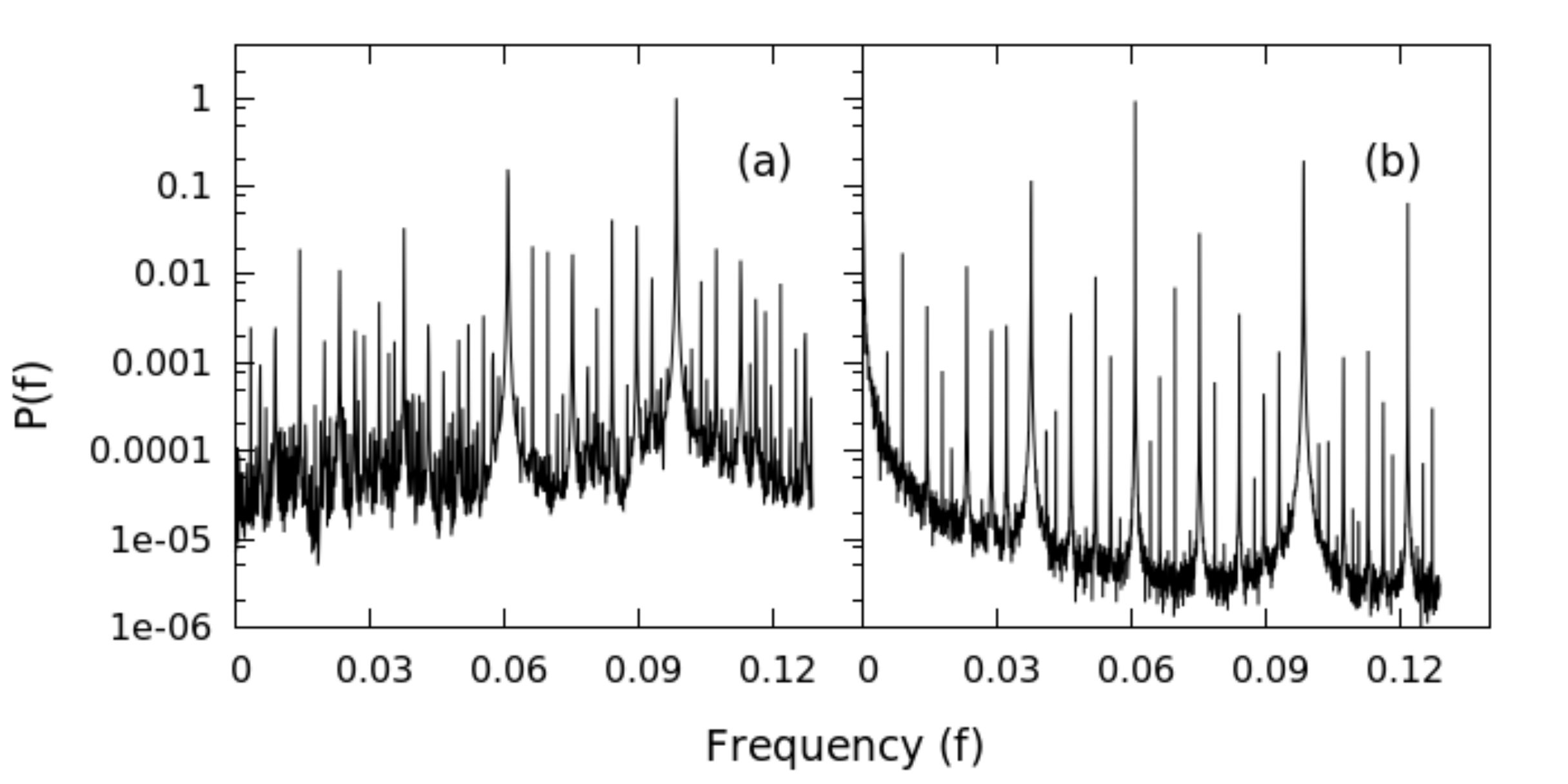}
\caption{\label{fig:strps}Power spectra for strobed time series of driven pendulum in (a) strange nonchaotic state evolved noiselessly and (b) quasi-periodic state with added red noise.}
\end{figure*}

\begin{figure*}
\centering
\includegraphics[width=.55\textwidth]{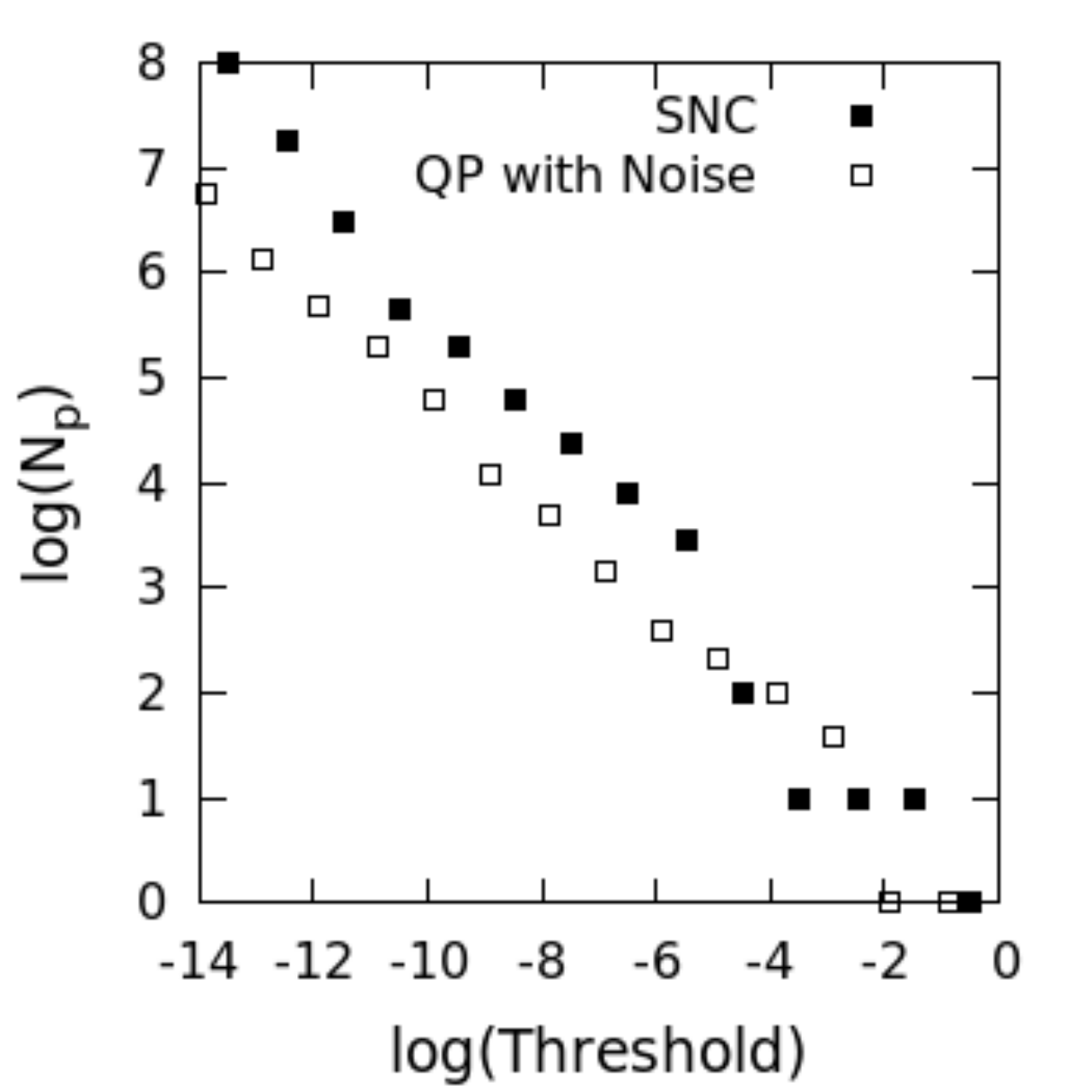}
\caption{\label{fig:ssqpsnc} Scaling behavior of the peaks in the strobed power spectrum for the doubly driven pendulum in regions of quasi-periodicity with added red noise and strange nonchaotic behavior without noise. The behavior is almost identical in both cases.}
\end{figure*}

To demonstrate this we add red noise with signal to noise ratio(ratio of standard deviations of signal to noise), 1.15 to the time series obtained from system.The time step for evolution is chosen as one tenth of the primary frequency, $F_1$ ($\frac{2\pi}{10\Omega_1}$). We strobe the $y$ variable signal(where $y=\frac{d^2(\phi)}{dt^2}$), along $F_1$ by picking every 10$^{th}$ point and then compute the power spectrum by dividing the time series into 32 segments. The power spectrum so computed is shown for all three cases in Figure \ref{fig:strps}. The scaling of the peaks is then calculated with increasing threshold power, after logarithmic binning to reduce statistical errors in the tail \cite{new05}. Here we define a peak as a local maximum in the power spectrum. This definition helps avoid coefficients that arise due to finite peak widths of the FFT. The log plots of number of peaks vs threshold power for the pendulum in the strange nonchaotic regime evolved in a noiseless environment and in the quasi-periodic regime with added red noise environment are shown in Figure \ref{fig:ssqpsnc}. Since the total number of peaks in the strobed noiseless quasi-periodic spectrum is rather small, the scaling remains distinctly different from these two cases, and is not shown here \cite{rom87}. The scaling behavior for both noisy quasi-periodic and strange nonchaotic behaviors is similar. If one is to rely solely on spectral scaling without any a priori knowledge of the system, one could be mistaken into believing that the noisy quasi-periodic data is derived from a strange nonchaotic state.
\begin{figure*}
\centering
\includegraphics[width=\textwidth]{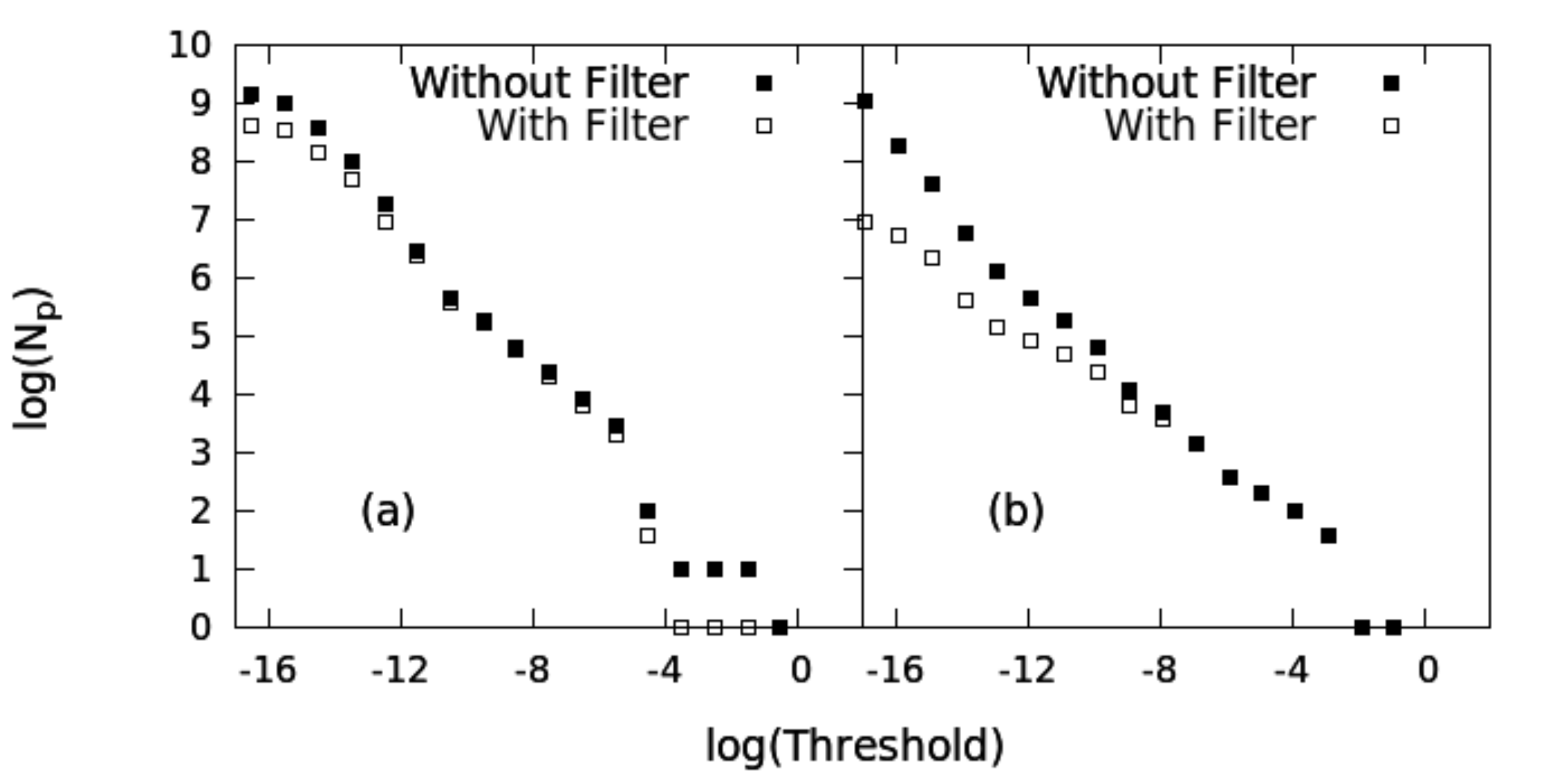}
\caption{\label{fig:ssqpwwo} Scaling behavior of the peaks in the power spectrum for the doubly driven pendulum. (a) is for the strange nonchaotic case without and with the bicoherence filter, (b) is for the quasi-periodicity with added red noise without and with the bicoherence filter. There is a marked difference in the scaling behavior once the bicoherence filter is applied for (b), while it remains identical for (a). This helps to identify the two states conclusively.}
\end{figure*}
We introduce a method of analysis using the main peak bicoherence defined above to detect the dynamics in such cases and check how many of these peaks are real. One of the main issues we see in using the previous definition of the main peak bicoherence function in this case, is the low Nyquist frequency of the strobed data. Hence if $f_m$ is the Nyquist frequency after strobing, and $F$ is the main peak in the power spectrum, we can only check for bicoherence upto $f_m$-$F$. This is not a problem for non strobed data, as most of the relevant frequencies lies below $f_m$-$F$. The main peak bicoherence function has significant values only for peaks of dynamical origin.  We exploit this observation to  filter the spectral scaling performed previously. In this filtering technique, we first count the number of peaks above a certain threshold. We then check how many of these peaks have a significant bicoherence. In cases where, $F$+$f$ is greater than $f_m$, we instead look for the significance of $b_F(F-f)$. We check the scaling behavior of this filtered number of peaks that have significant bicoherence values above the threshold. Then we see that the noisy quasi-periodic power spectrum shows a distinctly different scaling, where as the strange nonchaotic power spectrum continues to give the same scaling behavior. This is shown in Figure \ref{fig:ssqpwwo}. The filter reduces the total number of peaks in the noisy quasi-periodic power spectrum, by a factor more than 7. The bicoherence filter hence removes the ambiguity between noise contaminated quasi-periodic behavior and strange nonchaotic behavior.

We note that this technique is useful in deciphering hidden quasi-periodicity in the presence of additive white noise and when the system evolves in the presence of noise. The scaling with additive white noise is distinctly different from the scaling in the strange nonchaotic state. When the system evolves with noise, the power spectrum changes according to the equations of the system. In both cases however one can confirm the true scaling by applying a bicoherence filter.

\section{Bicoherence Analysis of Variable Stars}
\label{sec:sec4}
Nonlinear analysis has been used extensively in understanding the behavior of astrophysical objects \cite{mis06,zot13}. In this section we illustrate how the bicoherence techniques discussed above can be applied to understand the dynamical behavior of variable stars. We consider stars whose dynamical behavior has been analyzed in previous works using conventional methods \cite{sza10,lin15}. We concentrate on RRab Lyrae stars reported in the period doubled state and RRc Lyrae stars in strange nonchaotic state. Primarily we test our methods of analysis given in the previous sections in real world data. Second, we try to resolve the ambiguity between noisy quasi-periodic and strange nonchaotic behaviors by applying the filtering method suggested in Section \ref{sec:sec3}. Finally, we look forward to the new information supplied by the bicoherence function about the coupling between the different modes of the star. In the light of our analysis in Sections \ref{sec:sec2} and \ref{sec:sec3} it becomes important to analyze the bicoherence function to determine if the peaks in a power spectrum are of stochastic or dynamical origin which is imperative in the context of modeling these stars.  
\subsection{Stars in Period Doubled States}
\label{sec:sec4a}
\begin{figure*}
\centering
\includegraphics[width=\textwidth]{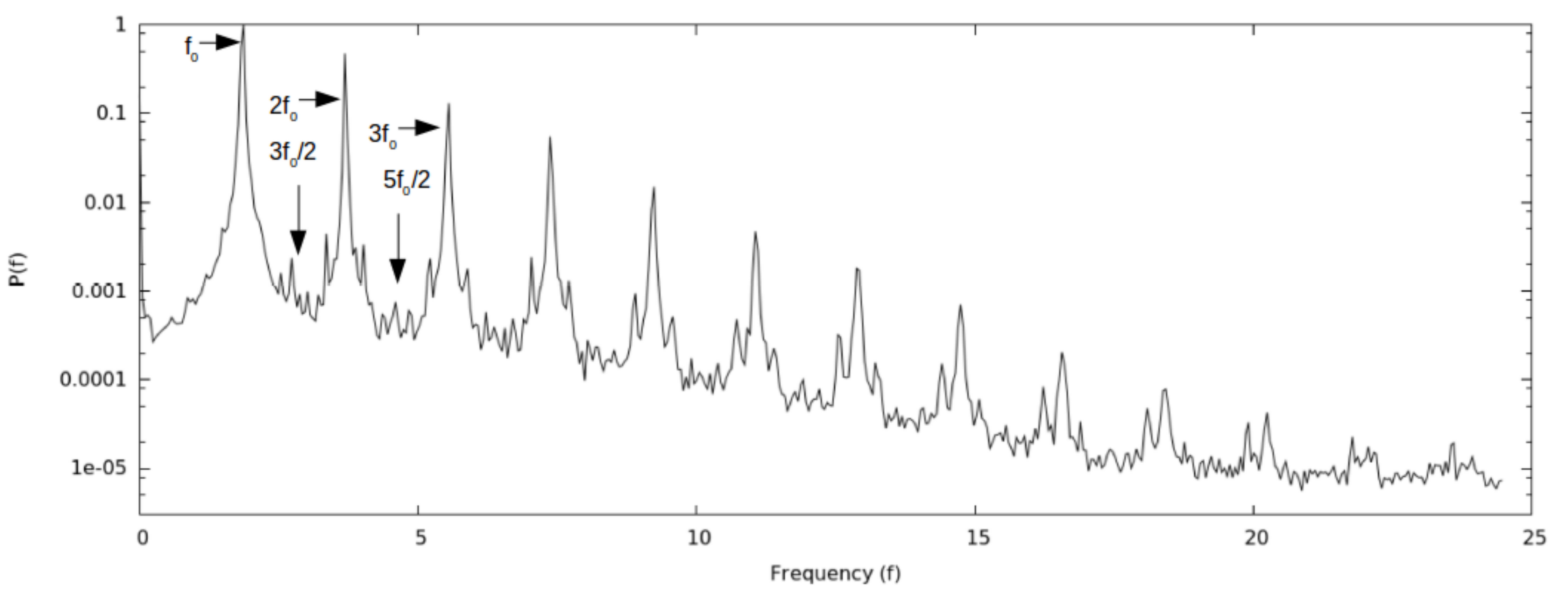}
\caption{\label{fig:pspds} Power spectrum for RR Lyrae star KIC 4484128. There are peaks at the primary frequency($f_0$=1.87 day$^{-1}$) and at integral multiples of $f_0$ and $f_0/2$. Minor peaks are seen between these, which may or may not be of dynamical origin.}
\end{figure*}

We initially present the analysis of the light curve of RRab Lyrae star  KIC 4484128. Due to gaps in the data, we consider segments of continuous data, find the power spectrum and average over the different segments. We use 22 segments of 1024 points each, sampled at .0204 days. (The frequency units are in day$^{-1}$.) The power spectrum (Figure \ref{fig:pspds}) shows peaks at the main frequency, 1.87 day$^{-1}$ and its integral multiples. The peak at half of the primary is overshadowed by the width the primary peak but the power spectrum shows peaks at 1.5 and 2.5 times the primary. It is hence evident that the star has undergone more than one period doubling bifurcation. Further we also see many other minor peaks in the power spectrum. As we mentioned in Section \ref{sec:sec2} one can find out if the minor peaks in the power spectrum are dynamically relevant only by a bicoherence analysis. This would help check if the less important peaks in the power spectrum are quadratically phase coupled. 
\begin{figure*}
\centering
\includegraphics[width=\textwidth]{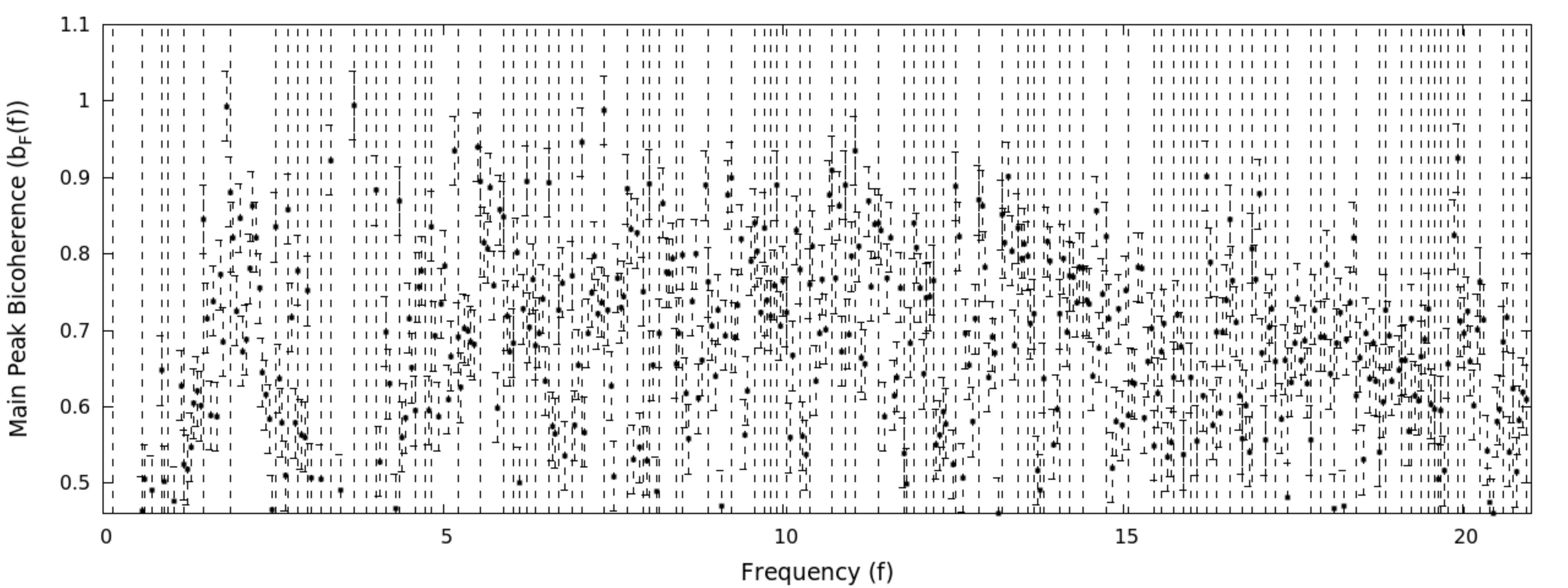}
\caption{\label{fig:mbcpds} Main peak bicoherence for RR Lyrae star KIC 4484128. There is a strong bicoherence at the primary frequency and at its integral multiples. The minor peaks in the power spectrum have significant bicoherence as well, suggesting they are of dynamical origin.}
\end{figure*}
\begin{figure*}
\centering
\includegraphics[width=.75\textwidth]{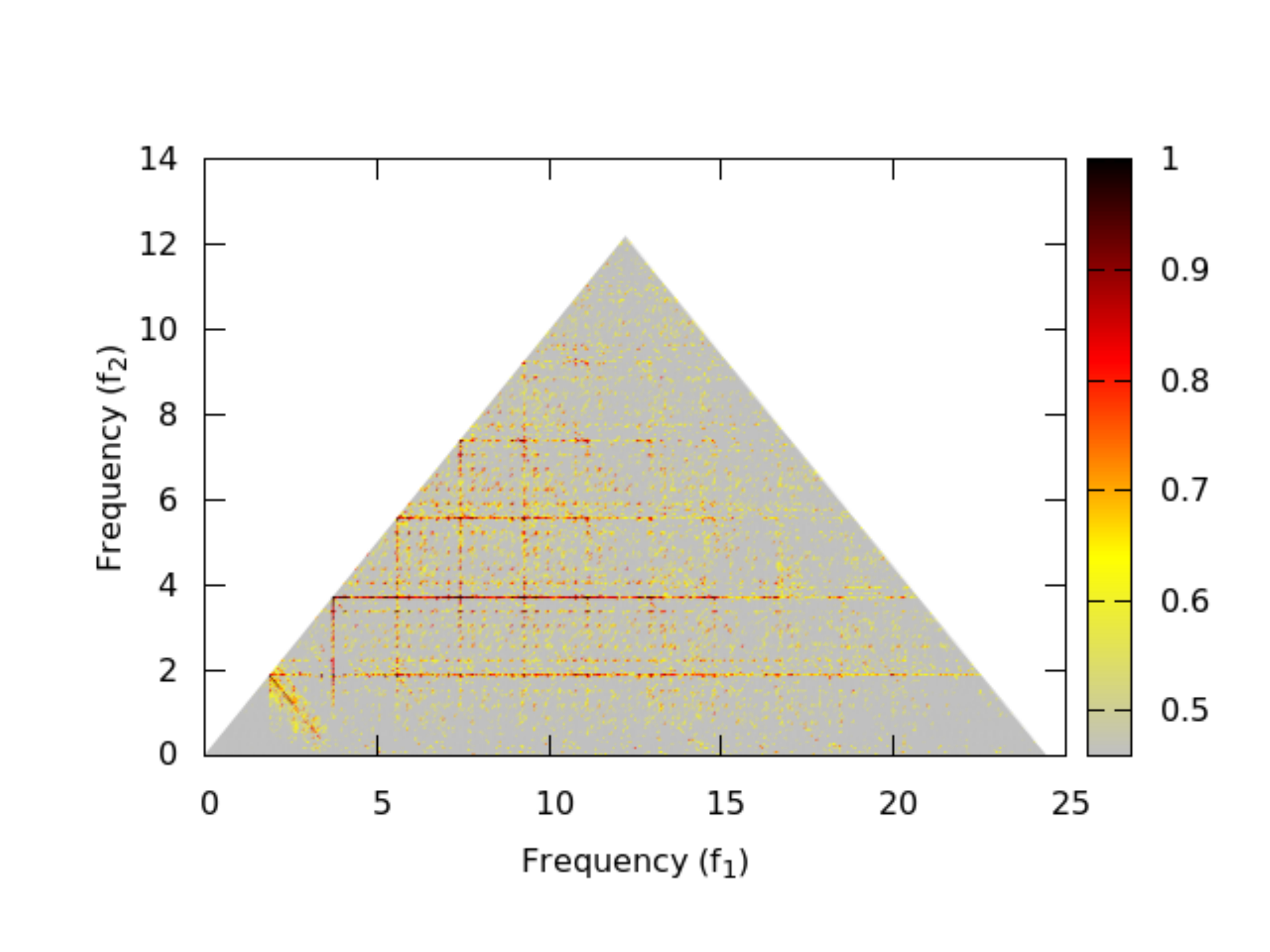}
\caption{\label{fig:fbcpds} Full bicoherence plane for RR Lyrae  KIC 4484128. Significant bicoherence can be seen at primary frequency $f_0$=1.87 day$^{-1}$ and at multiples of its period doubled frequencies. Continuous nature of this significance indicates dynamical origin for many in between frequencies also.}
\end{figure*}

Our aim is to establish if the inter-peak continuum in the power spectrum is relevant to the dynamics of the star or is indicative of some stochasticity in the system.  For this we present the main peak bicoherence in Figure \ref{fig:mbcpds}. It has a large number of frequencies showing significant main peak bicoherence, suggesting that the minor peaks in the power spectrum are also of dynamical origin.  The full bicoherence plane of the star, is shown in Figure \ref{fig:fbcpds}. We see high bicoherence at a large number of frequency pairs, in addition to period doubled ones. This suggests that KIC 4484128 is on the period doubling route to chaos. Similar analysis for RR Lyrae star KIC 7505345 yields qualitatively similar results. These stars hence have richer dynamics beyond the limit cycle behavior that can be confirmed as chaos using additional tests from fractal dimension calculations.

\subsection{Stars in strange nonchaotic state}
\label{sec:sec4b}
In this section, we look at the bicoherence function for two RRc Lyrae stars reported as having strange nonchaotic dynamics, namely KIC 4064484 and KIC 5520878 \cite{lin15}. In the light of our discussion in Section \ref{sec:sec3}, power law scaling alone does not guarantee strange nonchaotic dynamics. However using a bicoherence filter does successfully pick out true scaling behavior. Hence, we attempt to validate strange nonchaotic behavior in these stars, by looking at the main peak bicoherence function and subsequently re-validate the scaling behavior if need be. We mainly rely on the observation that power law scaling is observed for these stars even without strobing \cite{lin16}. We try to reproduce this power law scaling without a bicoherence filter. We restrain ourselves from strobing the data, as this would require interpolating between the different continuous segments. Earlier works suggest that interpolation may affect nonlinear quantification adversely \cite{geo15}.  

\begin{figure*}
\centering
\includegraphics[width=\textwidth]{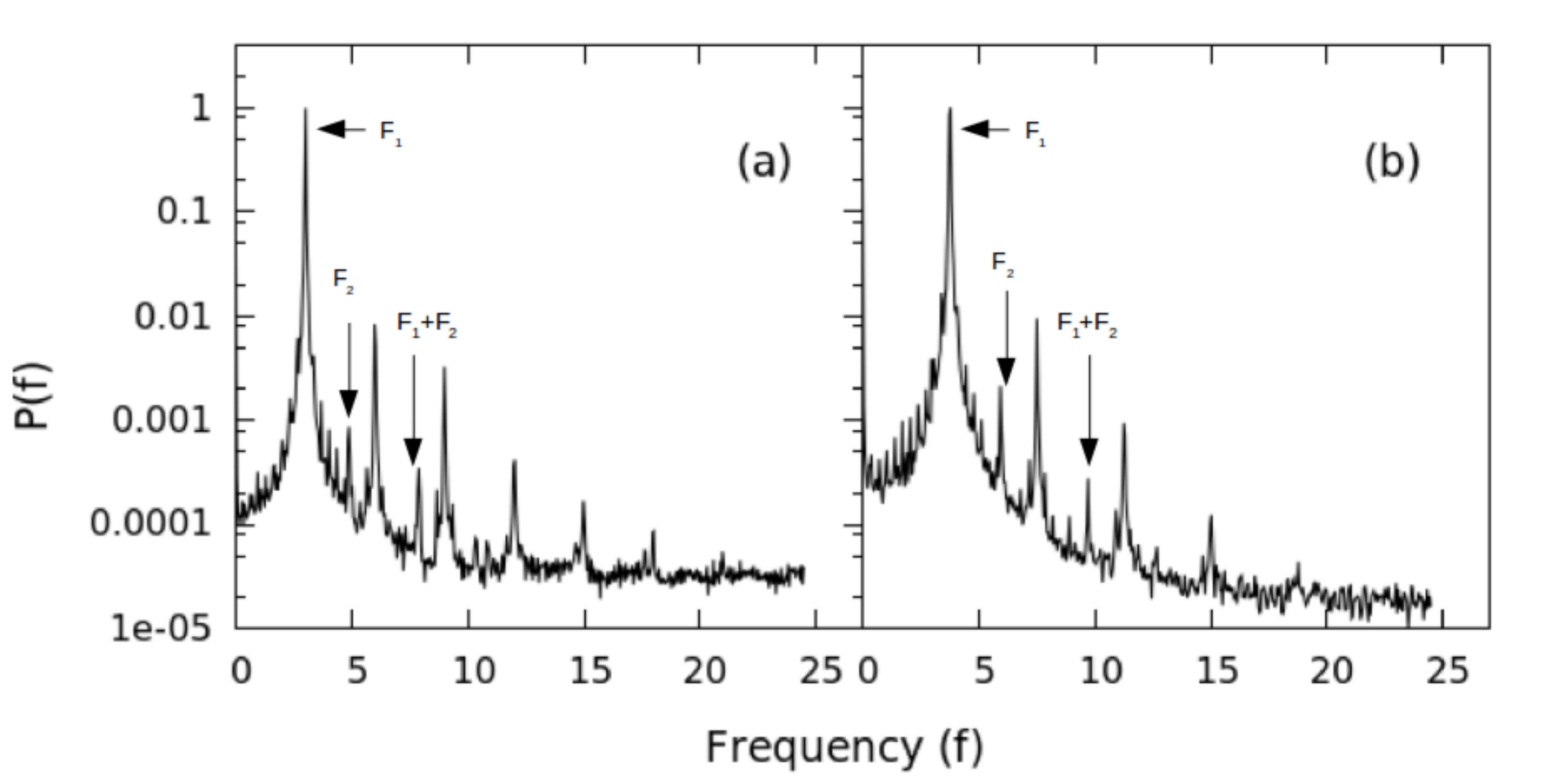}
\caption{\label{fig:pssncs} Power spectra for RR Lyrae stars (a)KIC 4064484 and (b)KIC 5520878. Major peaks can be seen at the primary frequencies ($F_1$=3.02 day$^{-1}$ and $F_2$=4.88 day$^{-1}$ for (a) and $F_1$=3.78 day$^{-1}$ and $F_2$=5.94 day$^{-1}$ for (b)) and their linear combinations characteristic of quasi-periodicity. Minor peaks seen between these, can be of stochastic or dynamical origin.}
\end{figure*}

\begin{figure*}
\centering
\begin{subfigure}{\textwidth}
\includegraphics[width=\textwidth]{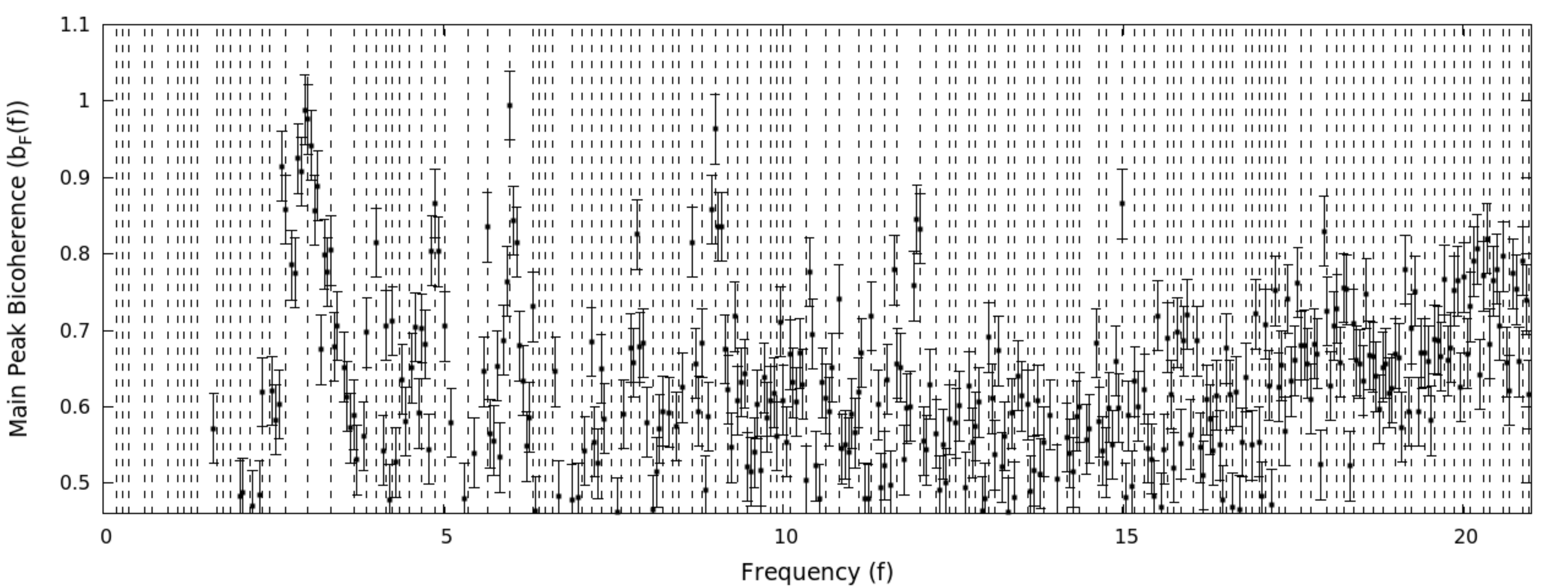}
\end{subfigure}
\begin{subfigure}{\textwidth}
\includegraphics[width=\textwidth]{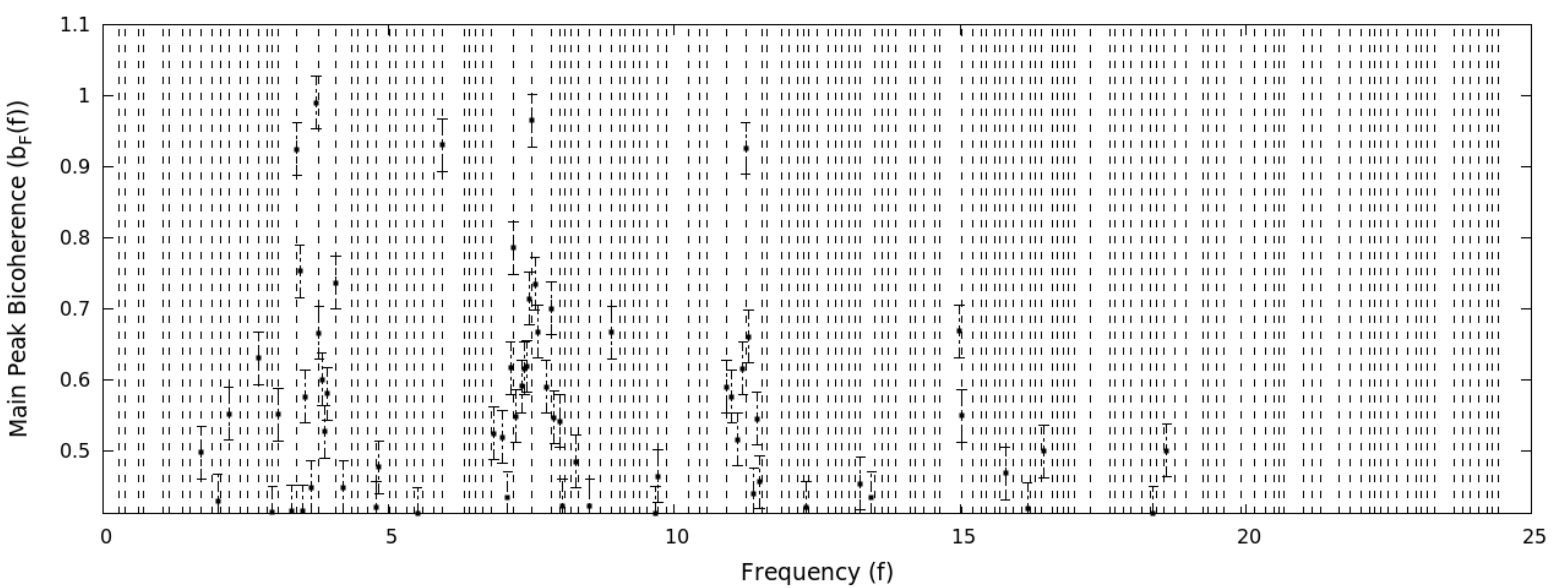}
\end{subfigure}
\caption{\label{fig:mbcsncs} Main peak bicoherence function for RR Lyrae stars (a)KIC 4064484 and (b)KIC 5520878.  The interpeak continuum can be confirmed as being due to the dynamics of the system for (a), as it has significant bicoherence for a large number of frequencies. However (b) does not show significant bicoherences for the intermediate peaks continuum.}
\end{figure*}

The power spectra for the stars are given in Figure \ref{fig:pssncs}. We see typical quasi-periodic behavior with peaks at the primary frequencies and their linear combinations. The primary and secondary peaks for KIC 4064484 are at 3.02 day$^{-1}$ and 4.88 day$^{-1}$; and are at 3.78 day$^{-1}$ and 5.94 day$^{-1}$ for KIC 5520878. The inter-peak continuum, can be indicative of dynamical noise, measurement noise, chaotic or strange nonchaotic behavior.
To distinguish between these possibilities, we take the main peak bicoherence of all four stars. This is shown in Figure \ref{fig:mbcsncs}.
We see that there is a significant bicoherence for the stars KIC 4064484 for a large number of frequencies implying that even the minor peaks in the power spectrum are due to the dynamics of the underlying equations. It seem likely, therefore that  it is indeed strange nonchaotic. We will consider the spectral scaling with a bicoherence filter to confirm this. For KIC 5520878  there is no significant bicoherence for the inter-peak continuum. Hence we may conclude that these peaks are not quadratically phase coupled to each other. The peaks may be due to the dynamical behavior of the underlying equations with a higher order coupling, or due to stochasticity present in the system or at the measurement end. For a preliminary test of the former, we consider the the tricoherence section at the primary frequency. No significant tricoherence was identified at the primary frequency, implying perhaps the minor peaks may indeed be of stochastic origin. 
\begin{figure*}
\centering
\includegraphics[width=\textwidth]{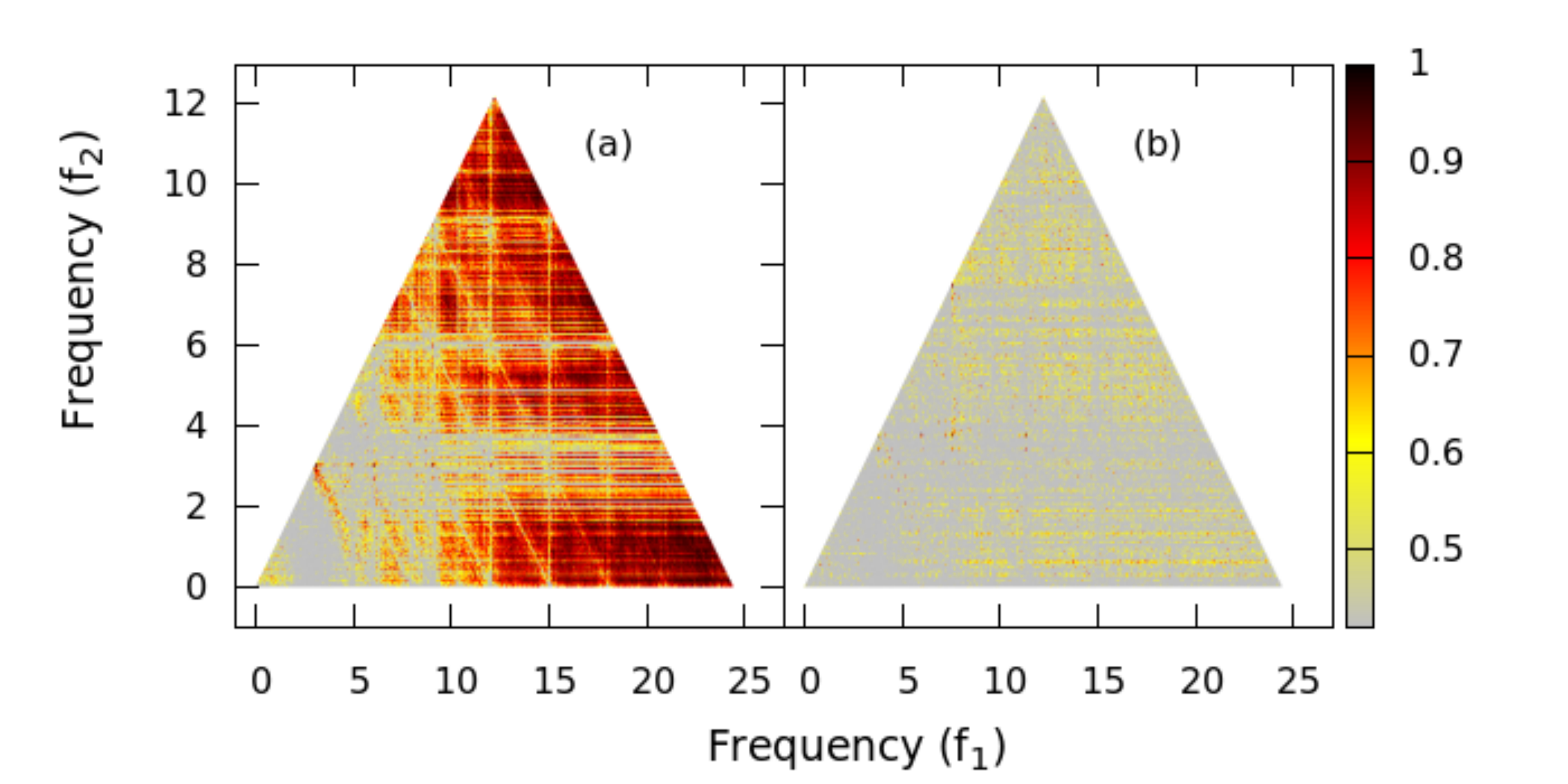}

\caption{\label{fig:fbcsncs} Full bicoherence planes for RR Lyrae stars (a)KIC 4064484 and (b)KIC 5520878. (a) has significant bicoherence across a large number of frequency pairs. (b) does not show high bicoherence, except at a few frequency pairs.}
\end{figure*}
We then look at the full bicoherence plane to confirm whether there exists other frequencies that are quadratically coupled to each other. The full bicoherence planes for both stars are shown in Figure \ref{fig:fbcsncs}. We find that as we deciphered from the main peak bicoherence function, the minor peaks of KIC 5520878 are not quadratically coupled to each other.

\begin{figure*}
\centering
\includegraphics[width=\textwidth]{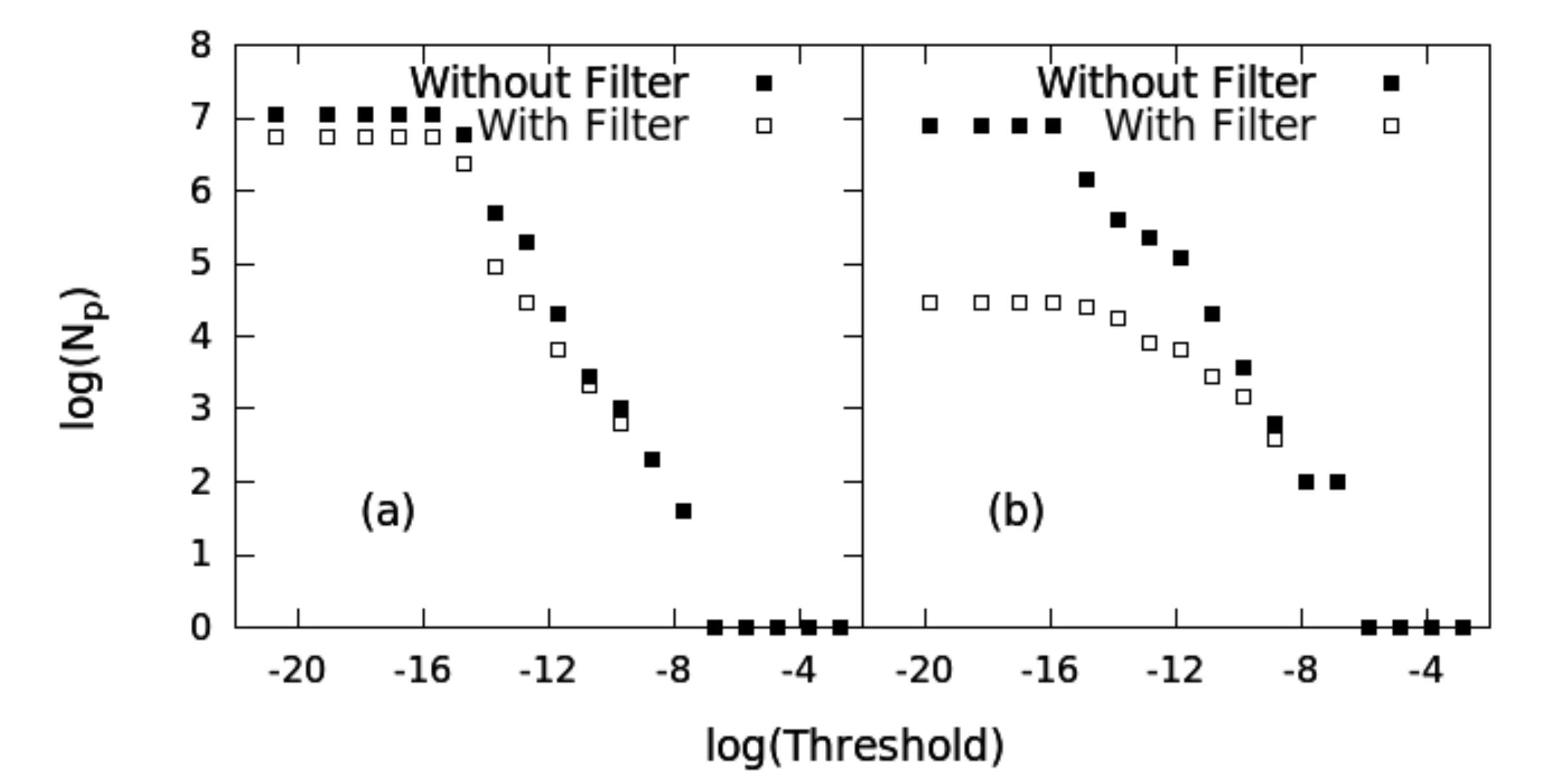}
\caption{\label{fig:ssstars} Spectral scaling for the stars (a) KIC 4064484 and (b) KIC 5520878. The scaling largely remains the same with and without the bicoherence filter for (a), validating the result that the peaks in the power spectrum are of dynamical origin and have a power law like scaling. The scaling behavior for (b) changes drastically, indicating that the star may not be exhibiting strange nonchaotic behavior.}
\end{figure*}
We now proceed to check how the scaling behavior for KIC 4064484 and KIC 5520878 change after the bicoherence filter is applied. This is shown in Figure \ref{fig:ssstars}.We see that the scaling behavior for KIC 4064484 is identical with and without the bicoherence filter, confirming that the peaks with significant bicoherence too have a power law like behavior. As expected, the scaling behavior for KIC 5520878 shows a marked change when the bicoherence filter is applied. This is shown in Figure \ref{fig:ssstars}. The persistent power law behavior in KIC 4064484 is indicative of strange nonchaotic behavior, as suggested by Linder et al. \cite{lin15,lin16}.
Similar analysis was also conducted on the stars KIC 8832417 and KIC 9453114, reported as exhibiting strange nonchaotic behavior \cite{lin15}. The former showed bicoherence similar to KIC 5520878, while the latter showed bicoherence and bicoherence filtered spectral scaling similar to KIC 4064484, suggesting strange nonchaotic behavior. Thus we conclude from our analysis that the underlying dynamics of the stars KIC 4064484 and KIC 9453114 differ from KIC 5520878 and KIC 8832417.

\section{Conclusions and Summary}
\label{sec:sec5}
We illustrate how the bicoherence function can be used to our advantage, for understanding the underlying dynamics of a system, especially in the presence of noise. The work presented is a study on the possibilities and methods of applying bicoherence function. We take the cases of limit cycle R{\"o}ssler, chaotic R{\"o}ssler, quasi-periodic driven pendulum and strange nonchaotic driven pendulum, as examples of standard systems to illustrate this. We also define the main peak bicoherence function and demonstrate its use as a quantifying technique for the latter two dynamical behaviors. 

In general for detecting the underlying dynamics of a system from which the time series was derived, we often rely on a battery of tests, starting from linear quantifiers like the auto-correlation and power spectrum,  to non-linear quantifiers like $D_2$ and multifractal spectrum. We advocate that this list is incomplete without the addition of the bicoherence function. We show that even when the system evolves in a stochastic background the bicoherence function picks out the related frequencies. This is especially useful in modeling a system, because the dynamical and stochastic parts are clearly distinguishable using the bicoherence function. In this context we claim that this method can bypass the conventional hypothesis testing using Fourier phase randomized surrogates, derived from data.
 
We also consider other dynamical states, like quasi-periodic and strange nonchaotic behaviors. Here we argue that the property of quasi-periodic systems to have peaks at linear combinations of the fundamental frequencies makes the bicoherence function a natural choice to examine them. We observe that the main peak bicoherence function will contain all the peaks that are present in the power spectrum. This is especially useful in the context of noise, when quantifiers like spectral scaling cannot clearly tell whether the data has been derived from a strange nonchaotic system or simply a noisy quasi-periodic one. In a real life scenario it is important to understand if the peaks in the power spectrum are indicative of chaotic or strange nonchaotic dynamics, or merely as a result of dynamical or measurement noise. We propose and successfully demonstrate the use of a bicoherence filter to pick out the true peaks in the power spectrum in the presence of noise, for the quasi-periodically forced pendulum.

As a useful and relevant application and to test our method on a real data set, we apply these techniques to the case of variable stars where period doubling and strange nonchaotic behavior were reported recently. We find the two stars KIC 4484128 and KIC 7505345, reported to have period doubling, exhibit many other frequencies that are of dynamical origin \cite{mos15}. This perhaps indicates that these stars are well into their period doubling route to chaos. We also consider the bicoherence for the reported strange nonchaotic stars  KIC 4064484, KIC 5520878, KIC 8832417 and KIC 9453114 \cite{lin15}. We confirm that the peaks in the power spectrum are indeed of non stochastic origin for KIC 4064484 and KIC 9453114. The scaling behaviors for these stars persist even after we apply a bicoherence filter, thus confirming their strange nonchaotic behavior. KIC 5520878 and KIC 8832417 however, show no significant bicoherence for the peaks in between the main peaks, and the scaling behavior changes drastically when a bicoherence filter is applied. Hence we can conclude that their underlying dynamics is quasiperiodic in origin. 

While the advantages of the bicoherence function in detecting the underlying non-linearity has been elaborated in the paper, it is imperative to note its limitations. The biggest, comes in the form the power of the dominant nonlinearity. It is well known that the bicoherence function will not characterize the system, if the power of the dominant nonlinearity is above 2. One can easily observe that if the power of the dominant nonlinearity is odd, the bicoherence function will be similar to that of noise. In such cases one has to look for higher order polyspectra like the trispectrum to characterize the system \cite{cha93}. Also the length of data required for a reliable estimate of the bicoherence function, is fairly large \cite{tot15}. Hence a bispectral analysis becomes impractical in cases where only short datasets are available. Further, the finite width of the peaks causes spread out regions of high bicoherence in the bicoherence plane. Another important drawback is the presence of data gaps in real-world data, which is known to have an adverse effect on the fast Fourier transform, while leaving quantifiers like D$_2$ largely unaffected \cite{geo15,mun16}. This in turn would greatly affect the calculation of the bicoherence. These practical limitations might have contributed to the relative unpopularity of the bicoherence function among dynamical systems' enthusiasts in the past. However, we have shown that the advantages offered by the bicoherence function far outweighs these issues, and should find its way into the pantheon of tests used to quantify chaos and nonlinearity in dynamical systems.



%
%

\end{document}